\documentclass[11pt]{article}

\usepackage{mathpazo}

\usepackage[margin=1in]{geometry}

\usepackage{hyperref}

\usepackage[defaultsans]{cantarell}
\usepackage[T1]{fontenc}

\usepackage[para]{footmisc}

\usepackage[small,compact]{titlesec}
\titlespacing{\subsection}{0pt}{*2}{*0.125}
\titleformat*{\subsection}{\sffamily\bfseries}
\titleformat*{\section}{\sffamily\bfseries}

\usepackage{fancyhdr}
\newenvironment{squishlist}                                                     
  {\begin{itemize}                                                              
    \addtolength{\itemsep}{-0.33\baselineskip}                                  
   }                                                                            
  { \end{itemize} }

\usepackage{tcolorbox}

\begin{document}

\thispagestyle{plain}

\mbox{ }\vspace{-0.7in}

\begin{center}
{\Large \sffamily \bfseries The Importance of Computation in Astronomy Education} \\
{
M. Zingale\footnote{Stony Brook University},
F.X.~Timmes\footnote{Arizona State University},
R.~Fisher\footnote{U Mass Dartmouth},
B.W.~O'Shea\footnote{Michigan State University}
}
\end{center}

\begin{tcolorbox}
{\sffamily \bfseries Executive Summary:} Computational skills are required
across all astronomy disciplines. 
Many students enter degree programs without sufficient skills
to solve computational problems in their core classes or contribute immediately to research. 
We recommend advocacy for computational literacy, familiarity with fundamental
software carpentry skills, and mastery of basic numerical methods by the
completion of an undergraduate degree in Astronomy.  

\vspace {0.1in} We recommend the AAS Education Task Force advocate for
a significant increase in computational literacy.

\vspace {0.1in} We encourage the AAS to modestly fund efforts aimed at
providing Open Education Resources (OER) that will significantly
impact computational literacy in astronomy education.




\end{tcolorbox}

\section{Computational Needs in Astronomy \& Astrophysics}

Computational skills are required at all levels of education
and research in astronomy.  Theoretical astrophysics is dominated by simulation instruments, 
often written in compiled and interpreted languages. 
Observational astronomy is entirely digital, with 
software pipelines for reducing and analyzing data.  Community tools, such as the Python-based
AstroPy\footnote{\url{http://www.astropy.org/}\\}, are actively developed for these pipelines.
The workflow in astronomy is often expressed in UNIX-like environments such as  OS X or Linux.
Students in secondary or undergraduate programs may be unfamiliar with
(and put off from) the command line and the job skills that it enables.

\section{Undergraduate Education}

Many astronomy and astrophysics programs encourage their majors to
take some computer programming classes.  For example, the State
University of New York transfer path for physics requires an Introduction to Computer Science 
in the first 2
years\footnote{\url{http://www.suny.edu/attend/get-started/transfer-students/suny-transfer-paths/pdf/transferSUNY_Physics.pdf}\hfill}.
However, this is where the encouragement of developing essential,
transferrable job skills in computation frequently ends. 

Astronomy students should be versed in elements of scientific computing and
basic numerical analysis in astronomy courses that leverage
community developed infrastructure. For example, Open Source web-based tools can 
be used to explore all stages of stellar evolution. Open data archives enable access to
galactic and extragalactic data.  These, and other, communuty
developments offer educators outstanding opportunities to bring
students directly into contact with real-world data, and to integrate
data analysis and computation into the curriculum. Some examples of
data-driven educational exercises include:
\begin {squishlist}
\item Inferring the mass, radius, and density of the historic transiting exoplanet HD209458b
\item Creation of a HR diagram from Tycho data
\item Examination of stellar interiors using MESA-web
\item Determination of the Hubble constant $H_0$ from Supernova Type Ia light curve data
\item Analysis of gravitational waves from the historic binary black hole merger GW150914
\end {squishlist}

We applaud the AAS's advocacy for increased literacy in scientific computation,
 as exemplified by the Hack Day events at recent AAS meetings.  We
suggest the AAS enhance its encouragement of sharing computational
tools, educational lessons, and projects amongst their members.

%

\section{Graduate Education}

A popular way to train graduate students in specialized codes and
techniques used in each astronomy discipline are summer schools
and workshops. We encourage the AAS to:

\begin{squishlist}

\item Extend and promote these training sessions
in association with the AAS meetings.  
For example, the Software Carpentry\footnote{\url{http://software-carpentry.org/}} 
sessions at recent AAS meetings are an excellent example of this training, 
and there is significant potential to further expand upon these sessions.  

\item Offer software instrument-specific training sessions
at the AAS meetings. For example, ``Best practices for CLOUDY\footnote{\url{http://trac.nublado.org}}'',
``Introduction to MAESTRO\footnote{\url{http://boxlib-codes.github.io/MAESTRO/}}'', or ``Advanced yt\footnote{\url{http://yt-project.org/}}'' can
provide critical training and community networking opportunities for graduate students.

\item Organize instructor training sessions for Software Carpentry,
to facilitate participants offering these workshops at their own institutions.

\end{squishlist}




\section{Open Source and Open Education Resources}

Open Education Resources (OER) are freely accessible, openly
licensed documents and media for teaching, learning,
assessing, and research. OER are among the leading trends in 
education, yet there is a paucity of quality material for
astronomy. A few notable exceptions include: (1)
astroEDU\footnote{\url{http://astroedu.iau.org}}, which
launched in February 2015, targets K-12 and is supported by the IAU
Office for Astronomy Development; (2) the Astrobetter
Wiki\footnote{\url{http://www.astrobetter.com/wiki/Wiki+Home}}, which
include links to user-contributed class slides, animations, texts, and
other resources;
(3) open-licensed texts such as the Open Astrophysics Bookshelf\footnote{\url{https://open-astrophysics-bookshelf.github.io}}
and others\footnote{\url{http://www.pa.msu.edu/~ebrown/lecture-notes.html}};
(4) MESA-Web\footnote{\url{http://mesa-web.asu.edu}}, a
web-based portal for stellar evolution aimed at secondary and
undergraduate education; and
(5) IPython/Jupyter notebooks for deployment of interactive computation-based exercises.

\begin{squishlist}
\item
We encourage the AAS to modestly fund efforts aimed at providing OER
material that will significantly enhance the use of computation in astronomy education.
\end{squishlist}

\section{Careers}

Computational skills and critical thinking are among the most
transferable job skills that an astronomy education can provide.
We encourage the continued advocacy by the AAS for increased
computational literacy, exemplified by the Hack Day events at recent AAS meetings, 
to provide skills that employers consistently seek.


\end{document}